# Unsupervised Multi-latent Space Reinforcement Learning Framework for Video Summarization in Ultrasound Imaging

Roshan P Mathews[1], Mahesh Raveendranatha Panicker[1], Abhilash R Hareendranathan[2], Yale Tung Chen[3], Jacob L Jaremko[2], Brian Buchanan[2], Kiran Vishnu Narayan[4], Kesavadas C[5] and, Greeta Mathews[6]

[1]Indian Institute of Technology, Palakkad, India
[2]University of Alberta, Alberta, Canada
[3]Hospital Universitario Puerta de Hierro, Majadahonda, Spain
[4]Government Medical College, Thiruvananthapuram, India
[5]Sree Chitra Tirunal Institute of Medical Sciences and Technology, Thiruvananthapuram, India
[6]Bhagwan Mahaveer Jain Hospital, Bangalore, India

## Abstract

The COVID19 pandemic has highlighted the need for a tool to speed up triage in ultrasound scans and provide clinicians with fast access to relevant information. The proposed video-summarization technique is a step in this direction that provides clinicians access to relevant key-frames from a given ultrasound scan (such as lung ultrasound) while reducing resource, storage, and bandwidth requirements. We propose a new unsupervised reinforcement learning (RL) framework with novel rewards that facilitates unsupervised learning avoiding tedious and impractical manual labelling for summarizing ultrasound videos to enhance its utility as a triage tool in the emergency department (ED) and for use in telemedicine.

Using an attention ensemble of encoders, the high dimensional image is projected into a low dimensional latent space in terms of: a) reduced distance with a normal or abnormal class (classifier encoder), b) following a topology of landmarks (segmentation encoder), and c) the distance or topology agnostic latent representation (convolutional autoencoders). The decoder is implemented using a bi-directional long-short term memory (Bi-LSTM) which utilizes the latent space representation from the encoder. Our new paradigm for video summarization is capable of delivering classification labels and segmentation of key landmarks for each of the summarized keyframes. Validation is performed on lung ultrasound (LUS) dataset, that typically represent potential use cases in telemedicine and ED triage acquired from different medical centers across geographies (India, Spain, and Canada). The proposed approach trained and tested on 126 LUS videos showed high agreement with the ground truth with an average precision of over 80% and an average F1 score of well over 44 $\pm$ 1.7 % (Mean $\pm$ SD) %. The approach resulted in an average reduction in storage space of 77% which could potentially ease bandwidth and storage requirements.

*All correspondence is to be addressed to Dr. Mahesh Raveendranatha Panicker; email: mahesh@iitpkd.ac.in*

## 1. INTRODUCTION

Ultrasound (US) has been one of the most promising clinical tools for the diagnosis and monitoring of various organs, particularly dynamic evaluation in cardiology and obstetrics and gynecology [1]. It is fast, safe and easily portable when compared to other imaging modalities like X-ray and Computed Tomography (CT). With increased portability and easy disinfection, ultrasound is well suited for point of care and in emergency medicine [2]. More recently, in the wake of the pandemic, the feasibility of using lung ultrasound (LUS) as a triage tool to detect lung involvement in COVID-19 patients is also being examined [3]. An unprecedentedly high number of cases during the peaks of the pandemic place a strain on clinical resources, highlighting the need for automatic classification of ultrasound data into normal and abnormal classes [4-6] as well as detection and segmentation of various critical signs to determine the severity of anomalies [7-10]. With rapid developments in deep learning, these tasks can now be automated [9, 10]. This could potentially allow lightly trained clinicians to acquire lung US and obtain automated expert interpretation in the absence of scarce human specialist expertise.

In recent years the use of ultrasound videos has also increased for examination of dynamic tissues like lungs. Assessment of findings such as pleural sliding and diaphragmatic movements helps detect pathology such as pneumonia or pneumothorax. Dynamic images are acquired using a 3D transducer or using a video loop acquired using the common 2D US transducer. Compared to single 2D US these videos cover a larger field-of-view and can be acquired accurately with minimal training [11] from remote locations. However deciphering key diagnostic characteristics among the myriad of general (and mostly redundant) information in the video is tedious both in terms of time and resources. Also, one of the important impediments faced today in transmitting the ultrasound videos in the case of telemedicine for an expert opinion and annotation is the large size of these US videos.

The clinical advantages of using ultrasound videos for triage in the emergency department as well as in the telemedicine applications have motivated this research and a cross functional multidisciplinary team has explored the ways and means of video summarization that eventually resulted in developing a novel approach that effectively summarizes ultrasound videos overcoming the inherent technical limitations through the use of unsupervised reinforced learning algorithms. The outcome of the proposed approach is twofold: a) reducing the size of the video, enabling telemedicine even in low resource settings and b) selection of non-redundant clinically relevant frames to construct the video summary, which is useful for both automated and human expert interpretation.

**1.1 Related work**

The research in ultrasound video summarization aims at improving the clinical outcomes while bounded by three constraints viz; a) a memory constraint to keep the number of frames in the summary as small as possible, b) the summarized video should not miss the diagnostically important frames and c) the selected frames should have minimum amount of redundancy. Related work in natural videos [12-16, 25] use approaches like hierarchical clustering of texture features [12] and block orthogonal matching [13]. Deep Learning techniques for video summarization include hierarchical recurrent neural networks [14] that use frame level features for video summarization. This consists of a tensor-train embedding layer and a hierarchical long short term memory (LSTM) layer. The tensor-train layer embeds video features (frame over time) into a lower dimensional space. The hierarchical LSTM has two layers, the first layer is a single LSTM which encodes the intra subshot temporal dependence and the second layer is a bidirectional LSTM which captures the inter subshot temporal dependence. A relation-aware assignment learning has been employed for the video summarization in [15]. The unsupervised approach presented in [9] exploits the clip to clip relations to learn relation-aware hard assignments by employing graph node concepts. The approach also employs a multi-task loss in terms of a reconstruction constraint and a contrastive constraint. In a recent work [16], reinforcement learning (RL) has been employed for video summarization. The video summarization problem is modeled as a deep summarization network (DSN) following the principles of RL [16]. The DSN predicts a key-frame likelihood probability for each video frame and then takes actions based on the probability distributions to select frames, forming video summaries. In [16], a novel reward function is also proposed which includes a diversity reward and a representative reward and the training is carried out. Since there were no ground truth labels employed for training, the approach falls under the unsupervised learning domain.

Unlike natural image videos, summarization of ultrasound videos is subjective. Selection of frames would typically be based on the presence of specific anatomical features and artifacts that are clinically relevant for diagnosis. For example, in LUS, the summary video should contain frames consisting of features described on LUS like A-lines, B-lines and pleural consolidations if present which are indicative of lung involvement (see Fig 1). Due to this rich diversity in image features and the presence of noise artifacts, feature based summarization techniques developed on natural images generally perform poorly on ultrasound videos. There has been hardly any work towards summarization of ultrasound videos except for the work in [17]. In [17], there is an attempt towards summarizing the videos from fetal ultrasound screening, employing the concept of RL proposed in [16]. The paper claims that the framework of RL is well suited for unsupervised video summarization which is a key requirement for medical videos. Major limitations of the above-mentioned video summarization approaches are the lack of contextual information and an attention mechanism that focuses on clinically relevant features. We propose a new unsupervised RL framework that incorporates this contextual information using an ensemble encoder and an improved attention mechanism in the decoder.

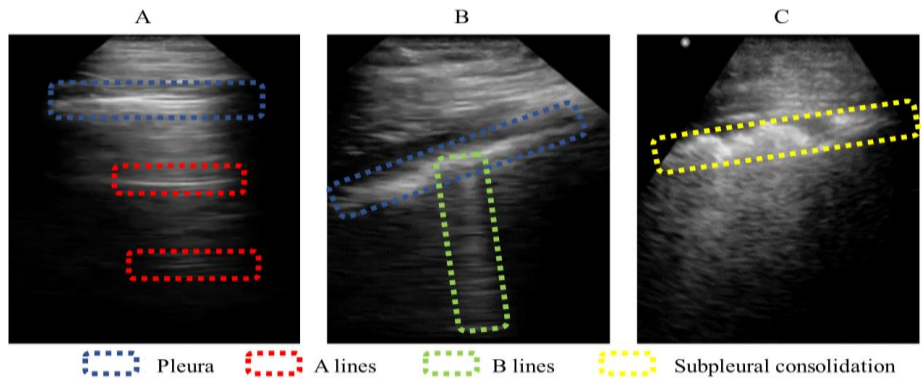

**Fig. 1 Manifestations in LUS - In image A, with A-lines (normal), in image B, with B-lines and subpleural consolidation in image C. Three or more B-lines and/or subpleural consolidation are pathologic and can indicate lung involvement in covid-19.**

### 1.2 Contributions

The key contributions in this paper are:

1. An exhaustive analysis of features from various encoders such as - a) an encoder of a classification unit which classifies a frame into normal vs. abnormal frame (e.g., presence of multiple B-lines or subpleural consolidations in lung is deemed an abnormal lung), b) an encoder of a segmentation unit such as a U-NET pretrained with manual segmentation images, c) an encoder of a convolutional autoencoder. Finally, an ensemble of encoders is proposed which encodes the frame onto different perspective manifolds - a classifier encoder which encodes the features responsible for abnormal lung such as B-lines or subpleural consolidations, a segmentation encoder which encodes the features corresponding to various biological landmarks in the given frame and an auto-encoder which provides the best compressed manifold representation of the frame. The ensemble of these encoded features are passed through a multi-head self-attention to highlight important sequences followed by an attention weighted averaging to generate a feature vector per frame.

2. An in-depth analysis of the decoders such as - a) a long-short term memory (LSTM) based decoder which has been the de facto architecture for temporal signals and b) a transformer based decoder which is effective in modelling long range dependencies.

3. Analysis of two probabilistic sampling schemes in the RL framework based on enforcing the selection of individual frames (simple Bernoulli scheme) vs. segment of continuous frames (a novel segment average Bernoulli) to generate the summary.

4. Two novel rewards for the RL framework are presented in this study - a) a structural similarity index matrix measure (SSIMM) based reward metric and b) a classification score based on lung state (healthy vs unhealthy).

5. A novel precision and recall calculation have been proposed which is based on similarity in the encoded latent space which offers lesser dependence on the exact temporal agreement between the summarised frames with the clinician annotated ground truth frames.

6. A unique video summary capable of enforcing only abnormal frames, overlaying the segmentations for easier triaging, and potentially used for compression in the spatial direction (bottleneck layers) in addition to the temporal direction (frame selection) for tele-medicine in low resource settings.

## 2. THEORY AND METHODS

Ultrasound video summarization highlights a fundamental limitation of supervised learning - the dependence on labeled data. Manual labeling of ultrasound videos for annotation of key frames relies on medical expertise and is tedious, expensive and time consuming. For example, a dataset of 1000 videos, each of 30 seconds duration, at 30 frames per seconds has 900,000 frames. With such a large number of frames, it becomes very time consuming for a clinician to annotate into sets of frames that are critical in the video for the purpose of summarizing it. Hence the most popular area of deep learning technique - the supervised deep learning is rendered ineffective in this particular application and necessitates to look for alternatives. The process of ultrasound video summarization involves the identification of key frames from a video which is typically done by a clinician who looks for certain biological landmarks. Similar to this, if we train an algorithm to pick out certain frames based on identification of certain landmarks we can try to mimic the behaviour of the clinicians doing the same. In the recent past, RL has gained huge traction in training algorithms to perform human-like tasks and with rewards that maximize the performance of the algorithm to human-like or even superior behaviours. Decoupling the rewards from the need for a human intervention (making them unsupervised) makes the whole process unsupervised. Hence, unsupervised RL can be used for the video summarization task. A generic framework for RL based video summarization is shown in Fig. 2. Unsupervised RL employs an agent that trains in such a way that it interacts with the environment to maximise its reward. The environment assigns a reward based on the result of that particular action on the environment. With this it is immediately clear that if we develop a method to reward the agent for selecting frames that are of relevance to the diagnosis, we can develop a model to summarize the given video.

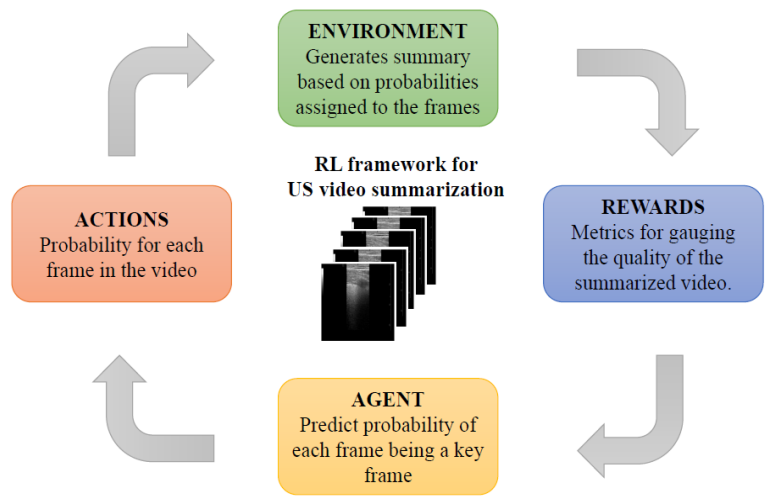

**Fig. 2 The adaptation of the RL framework for the proposed application of video summarization**

The RL framework consists of an agent, the RL algorithm whose task is to predict the probability of a frame being a keyframe and the assignment of these probabilities to each frame in the video is the action of the agent. Based on the actions of the agent, the environment produces a summarized video (by sampling the frames based on probabilities assigned by the agent) to which a reward or penalization is applied based on the reward metrics like how diverse and representative the summarized video is. The objective of the agent is to maximize the reward and thus indirectly maximize the metrics creating a summarized video.

A compact visual summary of the proposed video summarization framework is given in Fig 3. The proposed approach formulates ultrasound video summarization as sequential decision making based on the RL process. The essential building blocks of the approach are:

1. Encoder: Encodes single frame data into low dimensional feature space, which represents the spatial summarization in the case of a video.
2. Decoder: Use the features from the encoder to predict probabilities for each frame proportional to its importance to be selected into the summary set. The decoder ironically encodes the multi frame (temporal) data into the low dimensional feature space.
3. Summarization Algorithm: Determine the policy to select the frames into a summary set based on the probabilities output by the network.
4. Rewards: Rewards are metrics by which the decoder learns to generate probabilities which correspond to a better selection of frames into the summary set. Hence selection of appropriate rewards is important for the decoder to learn desirable video summarization.

The subsequent subsections will go over these blocks in detail describing their functionality in the overall framework.

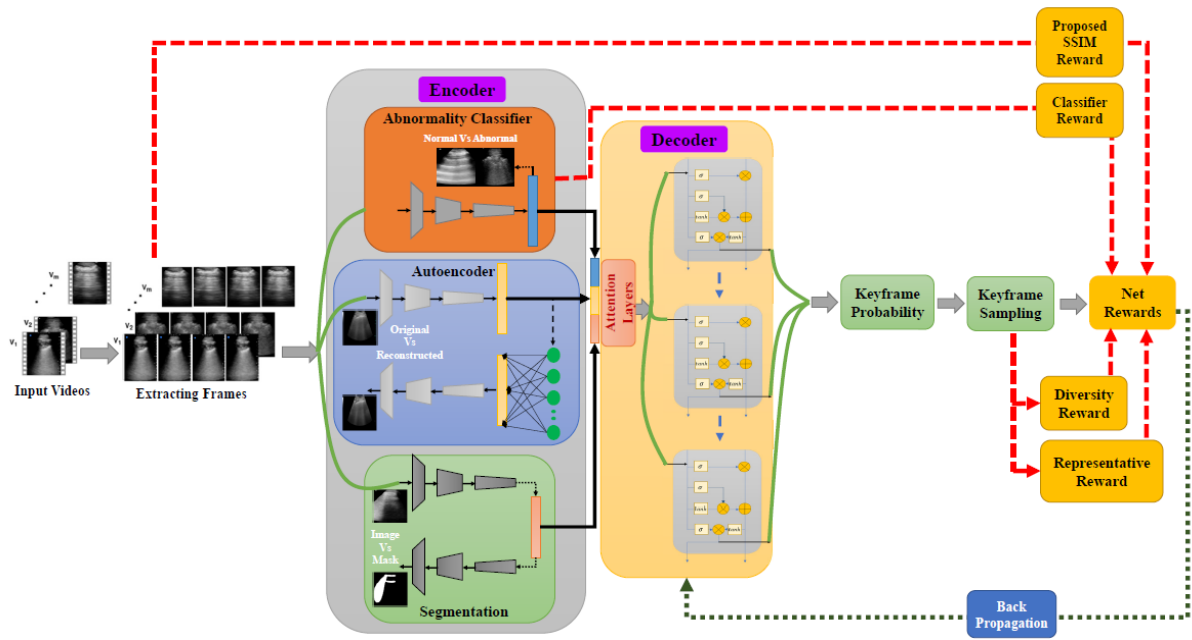

**Fig. 3 The proposed novel multi-latent space video summarization framework.**

## 2.1 Encoder

The encoder unit essentially consists of a spatial encoder which estimates the spatial latent vector representation of the given frame. These low dimensional spatial features of the frame are obtained by passing the frame through convolutional layers which represent the spatial information of the image frame into a feature map that can be understood by the summarization network. This is of special interest as different types of encoders capture different features in an image. In this work, a novel multi-latent space ensemble encoder is proposed which is the combination of a classifier encoder (trained with manual labels of the pathology), segmentation encoder (trained with manual segmentation labels) and an auto-encoder (with no manual labels). In each of these cases the latent space representation will be different and dominated by the application. A classifier encodes the features that are relevant in classifying an image into a certain category. An encoder trained at segmentation encodes features relevant to segment an image which typically are the important landmarks in the image. Finally, an autoencoder works to summarize the input image in the most compact fashion that can be reconstructed back to form the original image (hence doesn't depend on a certain distance or topology). Hence, it is readily seen that the choice of encoder plays a vital role in the selection of features from a given frame. A complete description of the various encoders in their individual capacity are given below.

*2.1.1 Classifier Encoder*

The design of a classifier encoder heavily relies on the application, for example in medical imaging, classifiers with shallow depths appear to perform better compared to deeper well-established networks employed for natural images [18]. For this reason, a shallow custom convolutional network is employed in this work, whose architecture is described in Fig. 4. The approach has four convolutional blocks which consist of 3x3 convolutional kernels, scaled exponential linear unit (SELU) and a down sampling by a factor of 2 using max-pool as shown in Fig. 4. Global average pooling is employed at the output of the fourth convolutional block to reduce the frame to a latent vector of 512 dimensions. The global average pooling helps to reduce the network parameters and thus prevents overfitting.

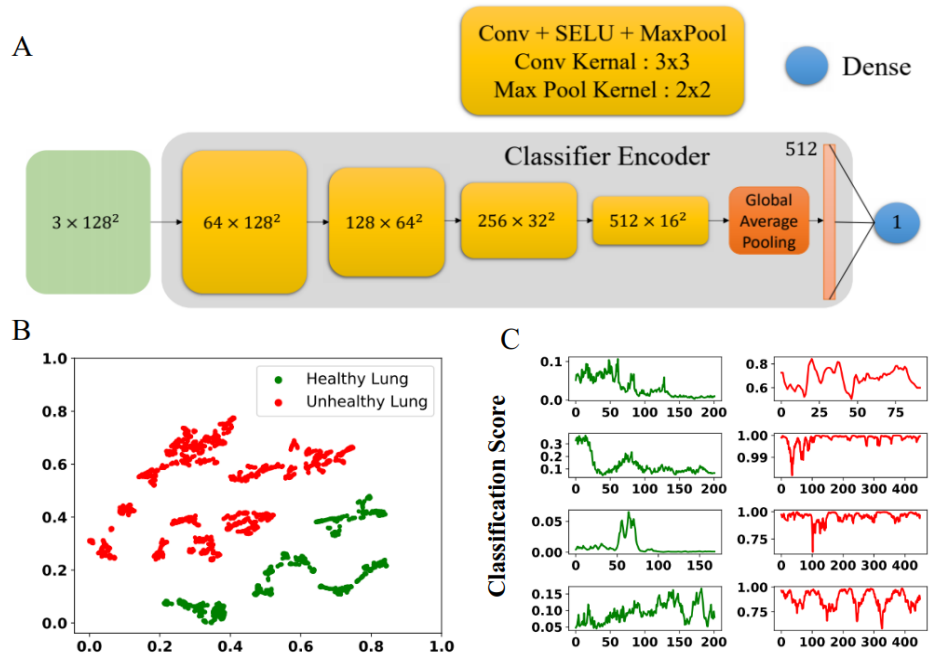

**Fig. 4 A) The classifier encoder architecture, B) the t-SNE measure of the classifier performance and C) classification score for LUS video frames for 4 healthy LUS and 4 unhealthy LUS videos.**

As discussed, the classifier is trained to detect abnormal lung vs. normal lung in LUS cases. The loss function is the typical categorical cross entropy given by (1)

$$E_{classifier} = -\sum_{i=1}^{2} y_i \log(\hat{y}_i) \qquad (1)$$

where $\hat{y}_i$ and $y_i$ are the predicted class and ground truth class respectively.

For training the classifier encoder, the LUS dataset consisted of 7428 images and it was tested on 2640 images with a classification accuracy of 95%. This ensures that the output of the global average pooling has spatial vector representation which could classify the given frame as normal and abnormal. This is extremely useful to produce a summary video biased towards abnormal frames which is of prime importance to the clinicians. The t-distributed stochastic neighbor embedding (t-SNE) of the bottleneck layer of the classifier as shown in Fig. 4 (B) clearly shows that the network has been trained to encode the image to feature efficiently. The classification score as predicted by the proposed classifier encoder for a few normal and abnormal LUS videos is shown in Fig. 4 (C), where ordinate shows the classifier score and abscissa shows the frame number. Any classifier score less than 0.5 is considered to be that of a healthy lung.

*2.1.2 Segmentation Encoder*

For the encoder based on segmentation task, the approach consists of identifying the relevant features extracted by the encoder of a pre-trained U-Net architecture [19]. The U-net architecture employed in the current work consists of a Resnet34 [20] encoder architecture and a 5 depth U-Net decoder as shown in Fig. 5. As an example, for the LUS video summarization, the U-Net based architecture is pretrained employing around 1000 images of segmentations of pleural line, A and B lines. The energy function is computed by a pixel-wise soft-max over the final feature map combined with the cross-entropy loss function as in (2).

$$E_{seg} = \sum_{x \in \Omega} w(x) \log(p_l(x)) \qquad (2)$$

where $x$ is the pixel position with $x \in \Omega$ with $\Omega \subset Z^2$, $w: \Omega \rightarrow R$ is an optional weight map to highlight certain pixels, $p_l(x)$ is the softmax function for each pixel with $l: \Omega \rightarrow \{1, \ldots, K\}$ is the true label of each pixel. It has to be noted that in this work, the number of classes is 1, which includes the combined segmented regions of pleura, A and B lines. Some of the results after the pretraining of the U-Net architecture is as shown in Fig. 5. It is evident that the U-Net is doing a good segmentation of the pleura and A, B lines and this ensures that the encoder part of the U-Net is pre trained to learn the features relevant to the segmentation of A and B lines.

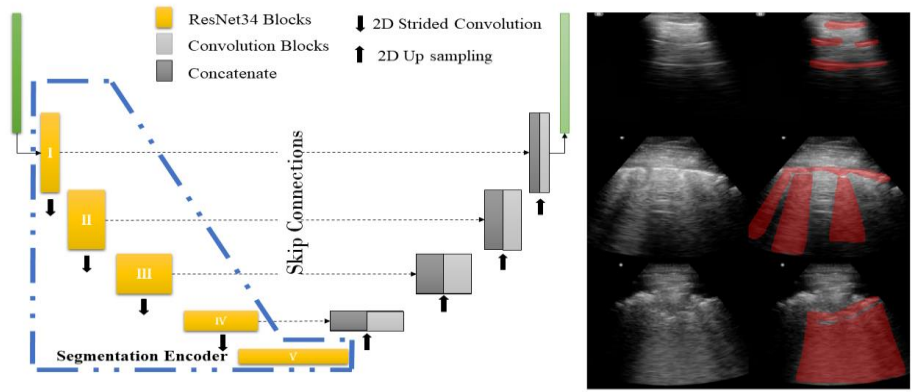

**Fig. 5 Architecture of the Segmentation Encoder (left) and segmentation results (right)**

The features extracted from the bottleneck layer of the U-net encoder (blue outlined block in Fig. 5) are used by the summarization network. It is proposed that the input image as well as the output mask generated by the U-Net be further fed back into the U-Net to reinforce anatomical structures in the input frame. Hence the final features to be used by the decoder are a weighted sum of features from the encoder of the input frame and the features of the segmentation mask as given in (3). This in fact will enforce the neural network to learn the features relevant to the segmentation mask.

$$Features = 0.7 * f\{I\} + 0.3 * f\{M\} \qquad (3)$$

where, $f$ is the segmentation encoder function, $Image\ (I)$ and $Mask\ (M)$ are the image and the respective reconstructed segmentation mask from the image. The weighting is done to ensure a larger portion of the features come from the input image, as the segmented mask maynot be 100% accurate and thus having a lower weight suppress the noise generated in the segmented mask. This at the same time reinforces the input anatomical features which are originally present in the video frame. To ensure the importance of (3), the t-SNE visualization of the features employing the relation in (3) and using just the image features extracted by the segmentation encoder are both estimated as shown in Fig. 6. The better region confinement of (3) is very evident from Fig. 6.

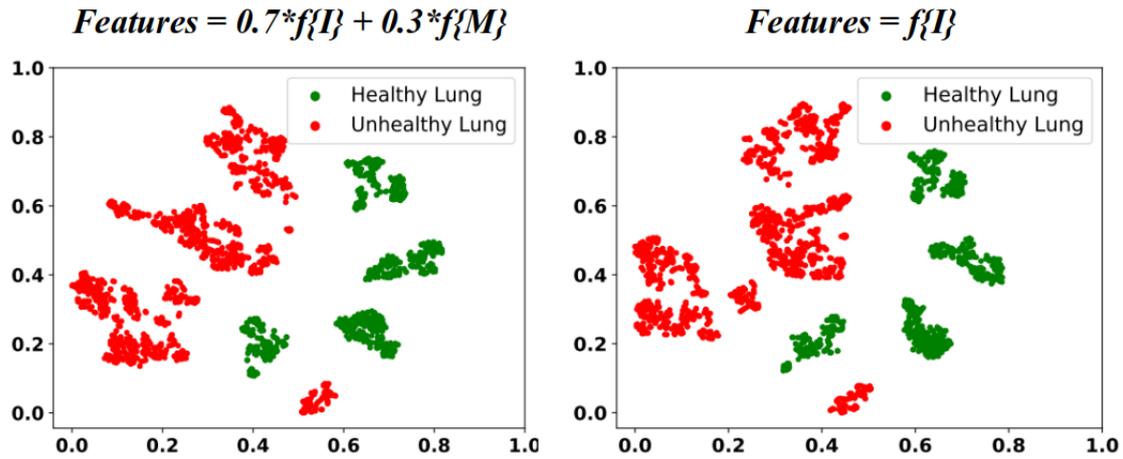

**Fig. 6 The t-SNE visualization for (3) (left) and by only employing $f\{I\}$ (right) for healthy and unhealthy lung states**

*2.1.3 Auto-Encoder*

Among the several variants of autoencoders, a convolutional autoencoder (CAE) [21] is used in this study. The CAE encodes the video frames to a latent space which is supposed to be the most compact spatial encoding of the frame. Thus, the input image is mapped onto a latent dimension using an encoder and the decoder is trained to reconstruct the image from this latent vector (although this is a lossy reconstruction). The architecture of the employed CAE is as shown in Fig. 7. The CAE is trained on 7428 images and it was tested on 1000 images with mean squared error (MSE) loss for the image reconstruction as in (4). An important point to be noted is that the training of such autoencoders does not require human intervention in the form of supervised labeling and thus is much easier to scale. Also from Fig. 7 B, it is clear from the t-SNE plot that the autoencoder has learnt meaningful compressed representations of the LUS and that it can possibly clearly delineate between healthy and unhealthy lung images.

$$E_{CAE} = \sum |x - \hat{x}|^2 \qquad (4)$$

where $x$ is the pixel position with $x \in \Omega$ with $\Omega \subset Z^2$ and $\hat{x}$ is the reconstruction of the $x$.

A few output samples of the CAE from the test set are given in Fig. 8. It is to be noted that the reconstruction is a lossy one (autoencoders are lossy) and the output from the autoencoder contains aspects from the input frame such as the presence of anatomical structures like the pleural lines (the white lines between the ribs), and B-lines (seen as white vertical lines/bands). Hence, although the output might not be an exact representation of the input, it has been able to encode features from the input that are diagnostically important.

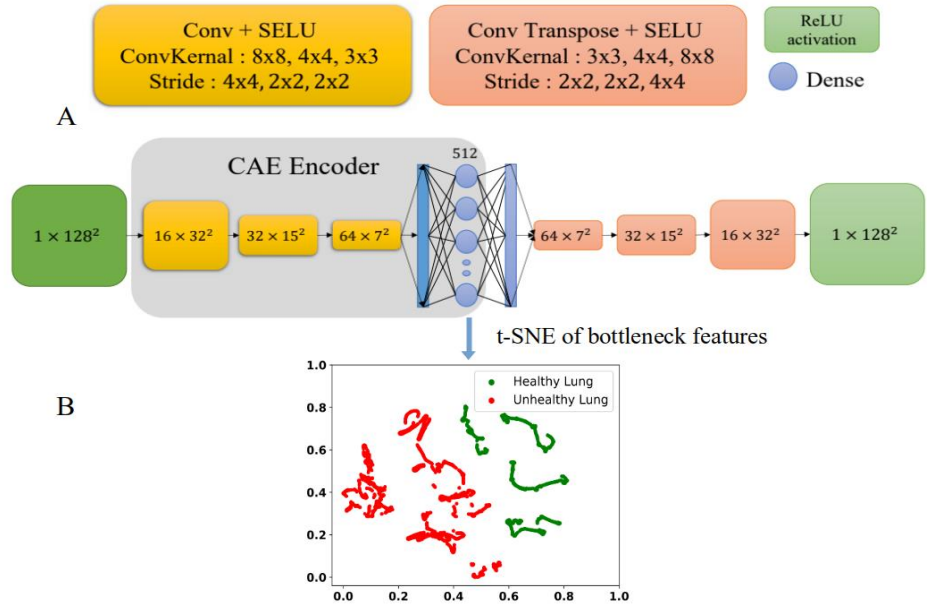

**Fig. 7 A) The architecture of the Convolutional Autoencoder and B) the t-SNE reduced visualization of the bottleneck features.**

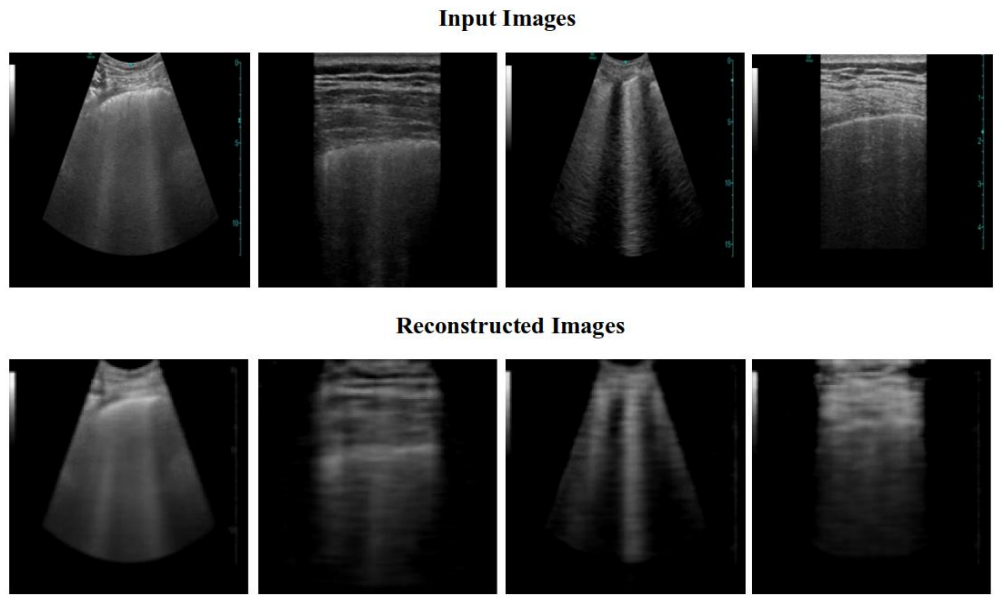

**Fig. 8 The first row are the inputs to the CAE and the second row the reconstructed images from the latent vector**

## *2.2 Proposed Ensemble Encoder Network*

From the above studies we have seen that each encoder encodes the same set of information in different ways. Hence it is natural to consider an ensemble encoder where the features are obtained from the ensemble of the encoders of the Classifier, Segmentation, and the CAE. In this section the details of the proposed ensemble encoder architecture are presented. All the encoder features are first normalized to keep them in similar magnitudes. Post normalization, the three encoder's features are input to a multi head attention (MHA) [22] to attend to the important features out of the three encoders. The output of the MHA undergoes a residual addition with the input sequence and followed by normalization to highlight important sequences among the encoders. The highlighted sequence is then attention weighted and summed to form a compact single vector representation of the three encoders. The architecture of the proposed attention network is shown in Fig. 9. It is to be noted that unlike the previous cases where the encoders were trained independently, the Ensemble Encoder is trained alongside the decoder by the rewards in a deep RL fashion to ensure that the features being attended to are tuned to aid the decoder in the summarizing task.

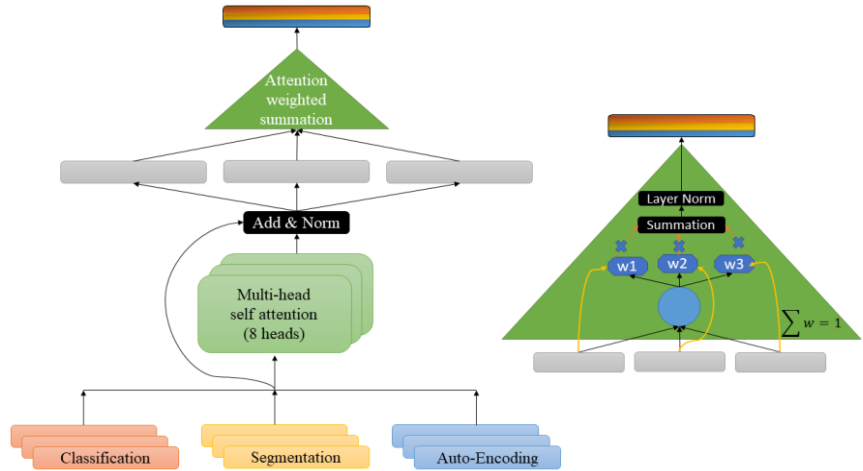

**Fig. 9 The proposed attention driven ensemble encoder**

### *2.3 Decoder*

The decoder in the summarization network predicts the probability for each frame's inclusion into the summary video with probabilities proportional to the frame's importance to be included in the summary set. Hence, each frame is assigned a probability between 0 and 1 with probability of 1 indicating the confirmed inclusion of the frame into the summary set and a probability of 0 indicating its exclusion. The input to the decoder are the features extracted from the encoder as detailed in the previous sections. Since the features extracted are from a set of frames in the video, they have a temporal orientation (order or sequence). So it is prudent that we use this information and use architectures that work with sequences. A major research focus in the area of sequence processing is the natural language processing (NLP) [23]. For years LSTM [23] has been the de facto architecture used for sequence processing tasks. Even for video summarization, all the previous related work in literature [12-17] uses LSTMs for video summarization. However the latest developments in NLP also use Transformers [22] (an architecture for sequence processing) for most sequence related tasks and have shown good promise. To the best of our knowledge, transformers haven't been used for video summarization tasks yet. The robustness of the decoders (LSTM and Transformers) for the task of LUS video summarization has been analyzed in detail in this work. The LSTM architecture employed consists of a bidirectional recurrent neural network (BiRNN) with a single hidden layer of size 256 coupled with a sigmoid activation. The transformer architecture is also a single layer feed-forward network using GeLU activation with a dimension size of 512 coupled with a 16 head multi-head attention. To prevent the feed-forward layer from overfitting, dropout regularizer is employed dropping one-fourth of the nodes randomly during training. Both the decoder architectures are topped with a fully connected layer with a sigmoid activation to output the per frame probability score. Unlike the LSTM where the video length is not a constraint due to its sequential input, the maximum frame number in a video is limited to 750 frames for the transformer architecture. A visual representation of the architectures of the decoder can be found in Fig. 10.

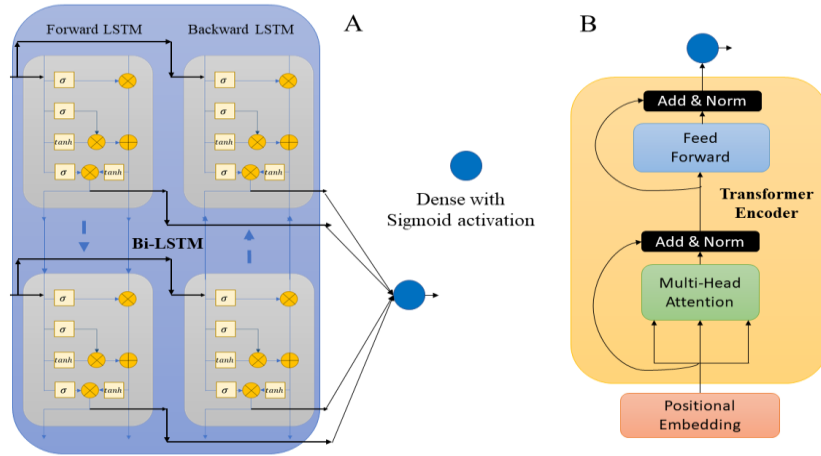

**Fig. 10 The two decoder architectures employed in this study, Fig 10 A represents the Bi-LSTM architecture and Fig10. B represents the Transformer (Transformer Encoder) architecture.**

## *2.4 Probabilistic sampling for the generation of summary set*

### *2.4.1 During Training*

The summarization algorithm uses the output probabilities by the decoder and generates the summary set, which forms the frames of the summarized video. Previous works [16, 17] in literature reports the use of Bernoulli sampling where each frame associated with a probability designated by the decoder network is sampled using a Bernoulli trial. The outcome of each Bernoulli trial is the weighted (based on decoder assigned probabilities) selection or rejection of a frame to be included in the summary set. The motivation for a Bernoulli sampling arises from the fact that when we introduce randomness into the system, it will promote learning in such a way that it minimizes the losses i.e. if a key frame has a probability $p$, then the learning algorithm would always try to increase $p$ such that the frame is selected a higher number of times in the Bernoulli trial. For the subsequent explanation of the proposed sampling schemes, let this approach be referred to as simple Bernoulli sampling (SBS) [16, 17]. Videos are a set of continuous frames and the SBS method disregards continuity of frames. To incorporate continuity in the frames being selected a novel sampling scheme is proposed where Bernoulli sampling is performed on segments/windows rather than each frame independently. To accomplish this, a moving average window is employed on a segment of five frames with a 1-lag overlap. The output of this is a segment averaged probability upon which we perform Bernoulli sampling to select segments of frames to be included in the summary set. Let this method be denoted as Segment Average Bernoulli (SAB).

To summarize, in this work ablation studies have been conducted between various sampling schemes as discussed.

1. *Simple Bernoulli Scheme (SBS):* The probabilities output by the summarization network for every frame is Bernoulli sampled. The frames that are selected in the Bernoulli trial are included in the summary.

2. *Segment Average Bernoulli (SAB):* This is a moving window average that computes the average probability of frames within a window. So, if there are $n$ frames, then for a window size of 5 frames we get $n - 5 + 1$ segment average probabilities. Upon this a Bernoulli trial is performed for each segment and the segments that are selected in the Bernoulli trial have all their frames included in the summary.

*2.4.2 During Inference*

During testing, unlike the case with training it is ideal to remove all randomness and thus Bernoulli sampling is the least favorable. One method to get consistent results with all inferences is to use the top 25% of frames by probability which are picked into the summary set. The downside of this is that there is no continuity inbuilt into such a sampling method and hence the summary videos might contain selection of single frames. By nature, videos are always a set of continuous frames and further in LUS it becomes a necessity to observe continuous movements of anatomical structures and hence it is proposed to use the moving window average to sample the top 15% segments by probability. Empirically it is seen that the top 15% segments by probability from the moving window average gives a summary length in between 20% to 25% which results in a compression by more than three-fourths. Let the selection of top 25% frames by probability be termed as T25 (Top 25%) and the selection of top 15% segments be termed as T15S (Top 15% Segments) for reference in subsequent sections.

*2.5 Rewards*

The decoder generated probability score for each frame in the video is used to sample the frames using a probabilistic sampling process as described in the previous subsection to form the summary video. The RL algorithm assesses the quality of the summarized video frames based on a reward function $R(S)$ as explained in detail below and the decoder is trained such that the rewards are maximized in the future iterations. Thus the decoder learns a policy which maximizes the reward and hence the quality of frame selection. In the case of generic video summarization [12-17], the most important reward metrics are the representative reward and the diversity reward. The representative reward is indicative of how well the summary video represents the whole video and can be formulated as a *k*-medoids minimization problem such that the mean of squared errors between video frames and their nearest medoids is minimal as in (5).

$$R_{rep} = exp(-\frac{1}{T}\sum_{t=1}^{T} min_{t'}||x - x'_t||_2) \qquad (5)$$

where T is the total number of frames in the given segment and t' is the medoid frame. While it is important to increase the number of frames in the keyframe, it is also important to have diverse frames in the summary and it is measured by the diversity reward as given in (6).

$$R_{div} = \frac{1}{|S|(|S|-1)} \sum_{t \in S} \sum_{i \in S, i \neq t} d(x_t, x_i) \qquad (6)$$

where $d(x_t, x_i)$ measures the cosine dissimilarity of two vectors. Similarity must be calculated between temporarily close frames and hence a threshold of 20 frames is set to ignore temporal similarity between distant frames.

While it is important to have a good trade-off between $R_{rep}$ and $R_{div}$ so that diverse and representative frames will form part of the key frames in the summary, at the same time, it is also important to introduce some rewards which are specific to the problem in hand. In the case of ultrasound imaging, clinicians look for abnormalities in the frame and hence it is important to have a bias in the selection of the frames towards larger number of frames being from the abnormal class. Towards this, we have introduced the classifier reward, $R_{Clsf}$.

$Frame\ Classification\ Score = x\ ;\ x \in [0, 1], 0\ healthy\ and\ 1\ unhealthy$
$$R_{Clsf} = \frac{1}{|S|} \sum_{i=1}^{|S|} x_i \qquad (7)$$

where Frame Classification Score is the classifier score as estimated by the classifier network and normalized across frames as in (7).

A novel structural similarity index matrix measure (SSIMM) is also proposed in this work, whose application is not restricted to ultrasound images. This score is designed to aid selection of diverse frames (dissimilar frames) as the keyframes based on the SSIM measure [24]. The various steps in the estimation of SSIMM based keyframes are as shown in Fig. 11. The first step is to create the SSIMM by calculating the SSIM between each of the frames in the video. The size of the SSIMM will be $N_f \times N_f$, where $N_f$ is the total number of frames in the video. Each row or column in the matrix (the matrix being symmetric) can be thought of as a measure of the SSIM of the given frame with all the other frames, i.e., the first column is the SSIM measure of the first frame in the video with all the other frames in the video including itself (the high intensity in the image corresponds to the self SSIM). In that case, it will be possible to have an absolute SSIM measure by collapsing the row or column of the SSIMM as shown in (8).

$$SSIMsig(x) = \sum_{y=1}^{N_f} SSIMM(x, y) \qquad (8)$$

where $x$, $y$ and $N_f$ are the respective row and column index and the total number of frames respectively. Each sample in the generated $SSIMsig$ corresponds to the sum of all the SSIM values of the frame identified by the index of the sample with all the other frames in the video, including itself. A closer analysis of the $SSIMsig$ reveals that, the higher values mean the frames are very similar to each other and the lower values mean they are very different. The transitions in the frame are where sudden dissimilarities exist (an example of this can be seen in Fig 11 around frame 60) and are of immediate interest to us being the frames which are diverse and qualify to be keyframes. To identify these sudden transitions, the next step is a sliding window slope estimation. For estimation of the slope, linear least-squares regression algorithm is employed with a window size of 10 frames. The window size is a hyper parameter which could be adjusted. The corresponding slope is as shown in Fig. 11 which clearly identifies the transition regions of interest to us. A simple thresholding of the same by using the mean of the slope signal as the threshold will give us the final key frames as can be seen by the bottom left plot in Fig. 11, where 1 stands for key frames and 0 stands for non-key frames.

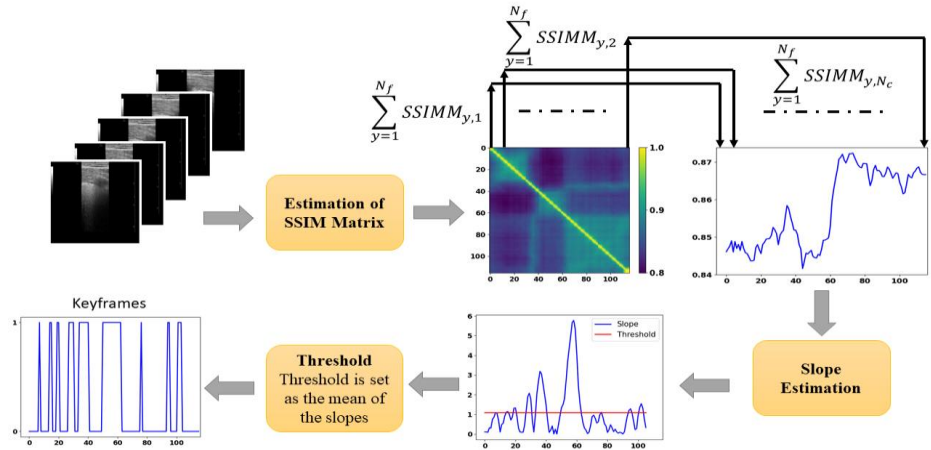

**Fig. 11 Proposed SSIM based reward measure**

The SSIM based rewards can be summarized as in (9)

$$R_{ssim} = \frac{1}{|S|} \sum_{i=1}^{|S|} SSIMsig(i) \qquad (9)$$

The final summary reward $R(S)$ is a weighted contribution from all the four reward metrics with the final value normalized between 0 and 1 as given by (10). Since, frames with abnormal state of the organ are usually preferred by clinicians, the classifier reward $R_{Clsf}$ has been weighted higher when compared to other rewards, however the weights could be employed as a clinician determined parameter.

$$R(S) = (2/3) * R_{Clsf} + (1/9) * R_{ssim} + (1/9) * R_{rep} + (1/9) * R_{div} \qquad (10)$$

### 2.6 RL Setup

*2.6.1 Optimization*

The training objective of the agent in the RL framework is to learn a policy $\pi_\theta$ with parameters $\theta$ to dictate the actions required to summarize the ultrasound video while maximizing the expected reward function $J(\theta)$.

$$J(\theta) = E_{p_\theta(a_{1\ to\ T})}[R(S)] \qquad (11)$$

where; $p_\theta(a_{1\ to\ T})$ is the probability distribution over the possible action sequences and R(S) is the summary reward.

A regularization that prevents the agent from picking too many frames to increase the reward is also employed in the optimization (12).

$$L_{reg} = \left\| \frac{1}{T}\sum_{t=1}^{T} p_t - \epsilon \right\|_2 \qquad (12)$$

where the $\epsilon$ determines the percentage of frames selected. Since we follow an unsupervised training, the rewards are used to calculate the performance of the agent and correspondingly the cost function is formed to train it. The overall cost function for the optimization for the parameters $\theta$ is then computed as $\beta L_{reg} - J$; where $\beta$ is a hyperparameter.

*2.6.2 Implementation*

The proposed video summarization approach was implemented in PyTorch using the open source repository for Deep reinforcement learning for unsupervised video summarization [16]. We trained the models on Tesla V100 Volta Graphical Processing Units hosted on the Compute Canada Cedar cluster. For the RL framework, the weight for regularization $\beta$ is set to 0.01, parameter $\epsilon$ that decides the percentage of frames to be selected as 0.5, the number of episodes to 10 and total epochs to 30. The Stochastic Gradient Descent with ADAM optimizer is used to train the models.

### 3. RESULTS AND DISCUSSION

The proposed framework has been validated on LUS dataset which includes 100 lung ultrasound videos taken from 40 subjects using different ultrasound machines (GE Venue, Philips Lumify, Butterfly network and Fujifilm Sonosite with 2MHz center frequency and Lung preset) and over various geographies (India, Spain and Canada) collected during the 2020-2021period. The data were collected in association with clinicians from Government Medical College, Kottayam, India, Sree Chitra Tirunal Institute of Medical Sciences and Technology, Thiruvananthapuram, India, Hospital Universitario Puerta de Hierro, Majadahonda, Spain and Department of Radiology and Diagnostic Imaging, University of Alberta.

The ultrasound exam was performed with the patient in supine or near-supine position for the anterior scanning, and in the sitting or lateral decubitus position for the posterior scanning. The probe was positioned obliquely, along the intercostal spaces. The LUS examination was obtained moving the probe along anatomical reference lines, 2nd-4th intercostal space (ICS) of parasternal, midclavicular, anterior axillary and midaxillary line (on the right side to the 5th ICS), whereas for the posterior chest, the paravertebral (2nd- 10th ICS), sub-scapular (7th- 10th ICS) and posterior axillary (2nd- 10th ICS) lines. A video clip was recorded along each anatomical line, recording at least 3 seconds at each ICS.

For testing, an independent set of 26 videos from 2021 are obtained and the keyframes are annotated by clinicians from Government Medical College, Thiruvananthapuram, India and Bhagwan Mahaveer Jain Hospital, Bangalore, India. The keyframes annotated by the clinicians are regarded as the ground truth and are employed for the testing and evaluation of the proposed approach.

*3.1 Evaluation Metrics*

As discussed in [17], precision, recall and F1-score are employed for measuring the performance of the proposed approach. The typical way to define precision, $P$, is as the temporal overlap between the summary (S) and the ground truth (GT) as $(S \cap GT)$ divided by the length of the ground truth ($\forall GT$) as given by (13). It is indicative of the proportion of ground truth present in the summary.

$$P = \frac{S \cap GT}{\forall GT} \qquad (13)$$

Recall, $R$, is defined as the temporal overlap between the summary and the ground truth as $(S \cap GT)$ divided by the length of the summary ($\forall S$) as in (14). It is indicative of the proportion of summary frames that are in the ground truth.

$$R = \frac{S \cap GT}{\forall S} \qquad (14)$$

Finally, the F1-score is given by the harmonic mean of the Precision and Recall,

$$F1 = \frac{2*P*R}{(P+R)} \qquad (15)$$

However, unlike natural videos, ultrasound videos which have a large number of frames convey the same diagnostic information and cannot be accurately summarized to a few frame indices. Hence instead of solely relying on the exact temporal location of the frames (frame index correlation), it is better to consider the feature level similarity of the frames. Towards this, a novel way to define the P and R in terms of the similarity at the feature level is proposed in this work as shown in Fig. 12. The first step of the approach is to reduce the dimension of the encoded feature vector to 2 from a size of $f$ (e.g., 512), which will aid in better feature similarity estimation using Cosine similarity measure. Assuming a total number of $N_f$ frames of feature size $f$ each, the t-SNE will create features $T_{xy}$ of size $N_f \times 2$. Let $GT$ be the ground truth annotations and $S$ be the summary estimated by the algorithm such as $GT, S \in \{0,1\}$ with a value 1 representing the presence of a key frame. The next step is to compute the Cosine similarity between each frame in $T_{xy}$ to $f_{GT}$, where $f_{GT}$ are the encoded features corresponding to the GT at the keyframe locations. To do so, every frame is compared against all frames in $GT$ and the Cosine similarity score $cosSim$ for a frame is the maximum of all the Cosine similarities. If a frame belongs to the summary set, $S$ and its similarity is greater than threshold, $Th$ (fixed as 0.999), then it is considered as true positive ($t_p$), else is considered as false positive ($f_p$). If a frame is not in the summary set, but its similarity is greater than threshold, $Th$, then it is considered as false negative ($f_n$). Employing the above measures, the precision (P) and recall (R) can be redefined as given by (16-17) and F1 score is given as in (15).

$$P = \frac{t_p}{t_p + f_p} \qquad (16)$$
$$R = \frac{t_p}{t_p + f_n} \qquad (17)$$

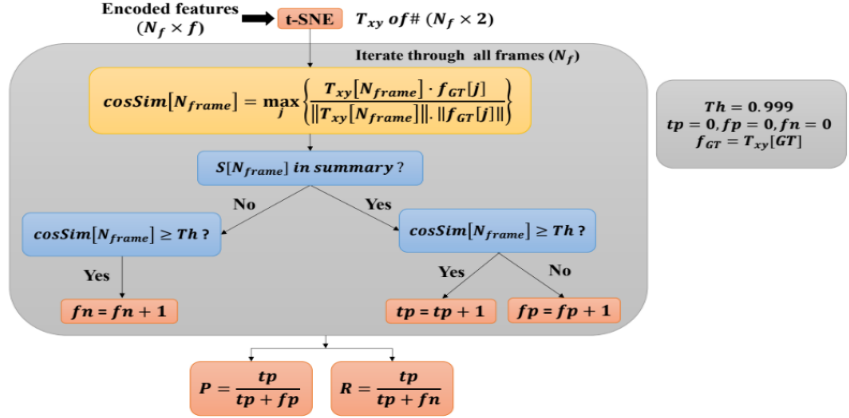

**Fig. 12 Proposed methodology for estimation of precision and recall giving importance to feature level similarity over exact temporal correlation with ground truth.**

In this work, since telemedicine is seen as a potential application, a new measure named reduction factor is also introduced and given by (18),

$$ReF = 100 * (1 - \frac{Summary\ length}{total\ length}) \quad (18)$$

Thus, higher the reduction factor, smaller will be the summary and easier for handling in the case of telemedicine.

*3.2 Ablation Studies employing Encoder-Decoder combinations*

The heart of the video summarization framework is the encoder-decoder combination and the choice of encoders (classifier vs segmentation vs CAE encoder) and decoders (LSTM vs transformers) play a significant role in the summarization process. In order to understand the effects of the same, each encoder in the above mentioned list is paired with the LSTM and transformer decoder to form the agent. The environment for the RL task consists of a Simple Bernoulli Scheme for generating the summary during the training phase. The choice of SBS as the sampling scheme to illustrate the effectiveness of the proposed methodology is due to its prior effectiveness tested in literature. During the inference phase the T15S is chosen to generate the summary. The results of the summarization are presented in Figs 13 and 14. The red plot is the clinician annotated ground truth, the green plot indicates the summarization by the LSTM decoder for the respective encoders and the blue plot indicates the summarization by the Transformer decoder. It is observed that the transformer's prediction for a single encoder decoder pair is almost similar for the three types of input encoders indicating robustness to the feature encoding process, whereas LSTMs display a wide variance in summary generated with the encoders used.

The quantitative results for the ablation studies mentioned above are given in Table 1. The values are averaged over five independent trained models to remove stochasticity from training the models. The inference from the table is clearly in favour of the transformer architecture for individual encoder cases which provides better inference for the given test dataset. Also, among the three encoders, the autoencoder trained encoder shows the best results in terms of a good trade off for the given inference dataset given that it follows a completely unsupervised training methodology. However, in the case of the proposed ensemble encoder, LSTM appears to be significantly better at accurately summarizing the video from the fused multi-latent state.

**Table 1. Quantitative Metrics for ablation studies involving encoders and decoders**

| Averaged over 5 independent trained models | | | | | |
|---|---|---|---|---|---|
| Encoder type | Decoder Type | Precision | Recall | F-Score | Reduction |
| Classifier | LSTM | 46.38 | 16.11 | 23.91 | 78.62 |
| | Transformer | 59.18 | 22.9 | 33.02 | 76.5 |
| Segmentation | LSTM | 61.58 | 24.87 | 35.43 | 75.08 |
| | Transformer | 63.04 | 23.95 | 34.71 | 76.3 |
| CAE | LSTM | 59.46 | 22.84 | 33 | 78.11 |
| | Transformer | 61.29 | 24.98 | 35.49 | 76.29 |
| **Attention Ensemble Encoders** | **LSTM** | **80.24** | **30.37** | **44.06** | **77.29** |
| | Transformer | 75.91 | 27.73 | 40.62 | 77.83 |

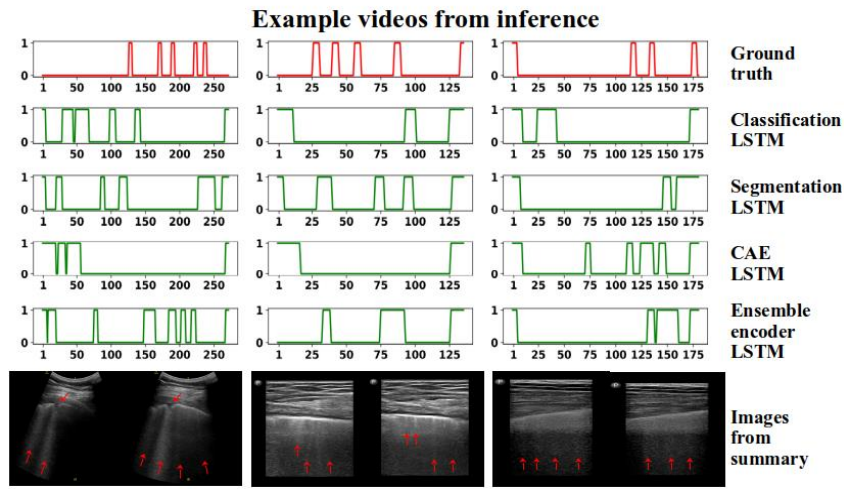

**Fig. 13 Summarization output for the features extracted from encoders by the LSTM**

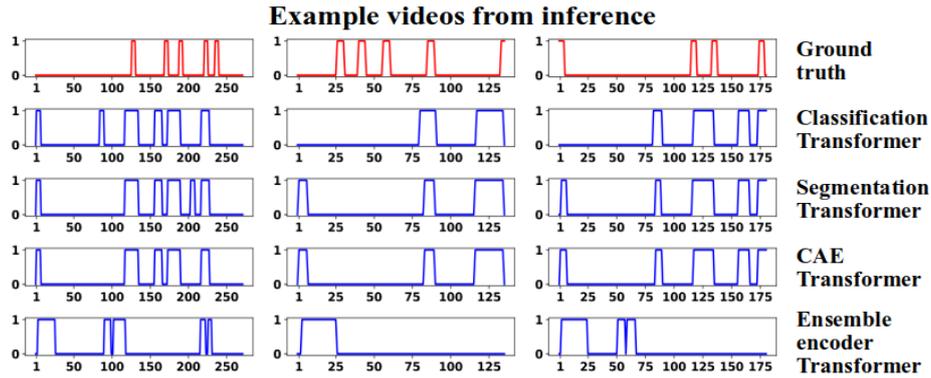

**Fig. 14 Summarization output for the features extracted from encoders by the Transformer**

A notable point is the apparent lower values of recall which is the expected behaviour due to the large number of false negatives resulting from the fact that the summary is restricted to a maximum length of 25% of the total frames in the video. This causes many frames with feature similarity to the ground truth higher than the threshold to be ignored in the process of summarizing. While increasing the summary length corresponds to fewer false negatives increasing the Recall and in process the F1 scores, it is not ideal as the aim of summarizing the video is to obtain an equivalent diagnostic information with the least possible summary length. Also, in medical imaging applications precision tends to precede other metrics in importance as it is crucial that relevant frames are not missed. The ensemble encoder - LSTM decoder pair provides very high Precision (over 80%) which are impressive results for an unsupervised training methodology further reinstating its importance.

### 3.3 Ablation Studies on sampling for summary generation

As discussed in the previous section, one of the key aspects of the RL scheme is the probabilistic sampling mechanism to generate the summary from the actions of the agent. For a systematic study of the effects of changing the environment in the RL framework, we keep all other variables fixed. Hence, the LSTM decoder is chosen as the agent for this set of the ablation studies due to its well established prior presence in literature. The main focus of this subsection is to investigate the effect of enforcing the selection of a segment of continuous frames over independent selection. Therefore the SBS during training is paired with T25 during inference and SAB during training is paired with T15S during inference. A quantitative analysis of the two methods during the training and inference phase are presented in Table 2. Again, the values are averaged over five independent trained models to remove stochasticity from training the models. It is seen that the SAB-T15S pair which enforces the selection of continuous frames provides better summarization for the classifier-LSTM and segmentation-LSTM pairs but, for the CAE-LSTM and Ensemble encoders-LSTM pairs, it performs subpar when compared to the SBS-T25 pair which have no restrictions on the selections of frames to form the summary set.

Since negligible difference is observed by enforcing the selection of continuous frames during training, the well established SBS is used during the training phase and T15S for inference for further ablation studies. The choice of T15S for inference over T25 is motivated from an application point of view where clinicians prefer a set of continuous frames over disjoint frames in the video.

**Table 2. Quantitative Metrics for ablation studies involving the two sampling schemes**

| Averaged over 5 independent trained models (not k fold) | | | | | |
|---|---|---|---|---|---|
| Enc-Dec type | Summary Generation | | Precision | Recall | F-Score |
| | Training | Testing | | | |
| Classifier - LSTM | SBS | T25 | 46.97 | 18.69 | 26.74 |
| | SAB | T15S | 52.30 | 18.14 | 26.94 |
| Segmentation - LSTM | SBS | T25 | 60.79 | 24.57 | 35 |
| | SAB | T15S | 62.71 | 24.76 | 35.50 |
| CAE - LSTM | SBS | T25 | 58.94 | 25.65 | 35.74 |
| | SAB | T15S | 58.95 | 23.03 | 33.12 |
| Ensemble ENcoders - LSTM | **SBS** | **T25** | **78.66** | **32.34** | **45.84** |
| | SAB | T15S | 75.39 | 28 | 40.83 |

## *3.4 Ablation Studies employing Rewards*

In RL, rewards are used to train the agent to pick actions that maximize the rewards. In this work, in addition to the typical representative and diversity rewards $R_{rep}$ and $R_{div}$, two novel rewards of classifier reward $R_{clf}$ and $R_{ssim}$ are introduced. More importantly, unlike [17], in this work, we propose a method which is fully unsupervised learning where we form rewards to summarize the network fully independent of a clinician. This is highly advantageous as there is a surplus amount of unlabeled data available and it also saves the clinician's time as annotating ground truth becomes a laborious task. During the training stage all rewards are normalized to be in the range 0-1 to prevent any reward from dominating the gradient updation of the decoder network. Fig. 15 displays the juxtaposition of the new rewards $R_{clf}$ and $R_{ssim}$ proposed in this study vs. the ground truth. We can see that the rewards and ground truth have a fair overlap (shaded in green) and thus the rewards used are ideal for training the summarization network.

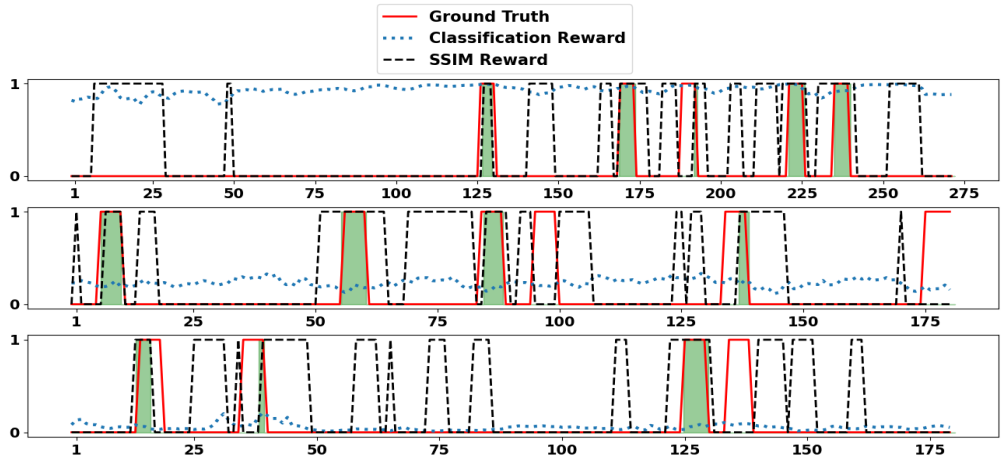

**Fig. 15 Juxtaposition of new rewards proposed vs. ground truth**

Further to quantitatively assess the importance of the rewards, ablation studies are performed where a particular pair of rewards is removed from training. To maintain a comparison with the results from using all the rewards during training (Table 1), the following section utilizes the SBS during training and T15S during inference. Table 3 gives the performance metrics for the ablation studies of rewards involved in training the network. Similar to the previous studies, the values are averaged over five independent trained models to remove stochasticity from training the models. The generic pair of rewards $R_{rep}$ and $R_{div}$ are equally weighted with a weight of 0.5 each and the pair of new rewards proposed in this study $R_{clsf}$ and $R_{ssim}$ are weighted as 2/3 and 1/3 each respectively.

**Table 3. Quantitative Metrics for ablation studies involving rewards in parts**

| Averaged over 5 independent trained models (not k fold) | | | | |
|---|---|---|---|---|
| Enc - Dec | Rewards | Precision | Recall | F-Score |
| Classifier - LSTM | Rep and Div | 46.93 | 16.26 | 24.15 |
|  | Clsf and SSIM | 49.35 | 17.17 | 25.48 |
| Segmentation - LSTM | Rep and Div | 60.78 | 24.20 | 34.62 |
|  | Clsf and SSIM | 59.88 | 23.89 | 34.15 |
| AutoEncoder - LSTM | Rep and Div | 58.19 | 22.30 | 32.24 |
|  | Clsf and SSIM | 60.81 | 23.57 | 33.97 |
| Ensemble Encoder - LSTM | Rep and Div | 75.19 | 27.57 | 40.35 |
|  | **Clsf and SSIM** | **77.33** | **29.30** | **42.50** |

An important detail from the results presented in Table 3 is that for a content specific summarization task as this study where the output is to summarize LUS video which are predominantly unhealthy lung sections (motivated by clinician's interest towards unhealthy lung sections), application specific rewards involving the classification score $R_{clsf}$ enhance the performance of the model in summarizing the video with frames that are diagnostically valued. This is supported from the fact that the pair of novel rewards proposed in this study $R_{clsf}$ and $R_{ssim}$ perform better than the pair of $R_{Rep}$ and $R_{Div}$ which are generic rewards.

Although the new pair of rewards $R_{clf}$ and $R_{ssim}$ proposed in this study proves to be of higher effectiveness than the generic pair of rewards $R_{Rep}$ and $R_{Div}$ for this dataset, it is fair to argue in favor of keeping all the rewards during the training stage as each of them cater to a different aspect during the summarization of the video and with rewards looking at diverse aspects the following framework can easily be extended to various other modalities of medical imaging.

In summary, this section has addressed the effect of different architectures in the video summarization pipeline and has presented a successful and robust video summarization architecture in the form of ensemble encoders with a LSTM decoder which produces over 44% F1-score and an 80% precision in ultrasound video summarization with the ability to compress the video by three-fourths its original size showing great promise in clinical applications, particularly telemedicine. It is remarkable that the above performance has been achieved using an unsupervised training framework on a diverse dataset. Unlike other imaging modalities, ultrasound scans are highly dependent on the individual clinician and thus videos from different US machines (different presets, mixture of scans using curved and linear probes) from different countries (different clinicians following independent scanning techniques) show wide variation in scans. Despite these differences our proposed framework was able to provide superior summarization indicating a possibility of even greater performance to standardized datasets which involve a single type of US machine with predefined presets and uniform scanning techniques by clinicians.

## 4. CONCLUSIONS

This study focuses on a resource-efficient ultrasound video summarisation pipeline, which can use unsupervised learning. This approach has vital relevance in a) critical clinical care situations where expert image interpretation is needed but human expertise is scarce (e.g., during the ongoing COVID-19 pandemic), and b) telemedicine and point-of-care medicine, where decreasing transmission bandwidth can remove limits on the rate at which videos are sent for telemedicine consultation (either to human experts or cloud-based artificial intelligence). The work demonstrated herein provides promising results validated with extensive clinical data with our methodology yielding Precision and F1-scores over 80% and 44% respectively with excellent video compression rates well over 77%. Also, the methodology described here is not limited to lung ultrasound and can be further generalized to any ultrasound video, with many clinical applications, such as in echocardiography, or in any type of ultrasound study in which a novice user may have unknowingly obtained some good-quality images within a lengthy video. The proposed work has the capability of highlighting the clinically relevant frames showing key abnormalities and landmarks.

The present work demonstrates a prototype for a robust ultrasound video summarization pipeline and future work would include analyzing the proposed system with well pruned US scans which are standardized with respect to probe geometries - either curved or linear, focus on the lung (avoiding extensive rib shadowing and artifacts or other organs) and ultrasound machines with specific preset settings for image acquisition etc. On the algorithmic front, introduction of novel rewards for the identification of specific biological landmarks for preferably selecting frames which are diagnostically valid frames rather than artifacts will also be addressed.

## ACKNOWLEDGMENTS

*The authors would like to acknowledge the funding from the Department of Science and Technology - Science and Engineering Research Board (DSTSERB (CVD/2020/000221)) for the CRG COVID19 funding. The authors are grateful to Compute Canada and the NVIDIA and CDAC for giving access to the PARAM SIDDHI AI system for providing the computing resources for the project.*

# Supplementary text for
# Unsupervised multi-latent space reinforcement learning framework for video summarization in ultrasound imaging

Roshan P Mathews, Mahesh Raveendranatha Panicker, Abhilash R Hareendranathan, Yale Tung Chen, Jacob L Jaremko, Brian Buchanan, Kiran Vishnu Narayan, Kesavadas C and, Greeta Mathews

August 2021

## Code and Data Availability

As supplement material to the main text, the code for generating the summary video, the prototype model and a few examples of lung ultrasound scans and their summaries generated by the summarization algorithm are provided in the GitHub repository [1] as additional material for reference.

## Preferential Summarization of Key Landmarks

To demonstrate the working of the video summarization algorithm an example video from the given dataset is used to highlight the visual differences between the complete ultrasound video and its summary (refer Supplementary Fig.1 and video name SupVid4 and its summary in [1]). It can be seen that the algorithm for summarizing the video has preferentially selected larger segments of frames with B-lines over frames that don't. Also, it is worth noting that the complete video has about 450 frames whereas the summarized video is just under 100 frames giving a compression ratio over three-fourths the full video length whilst preserving diagnostically important information.

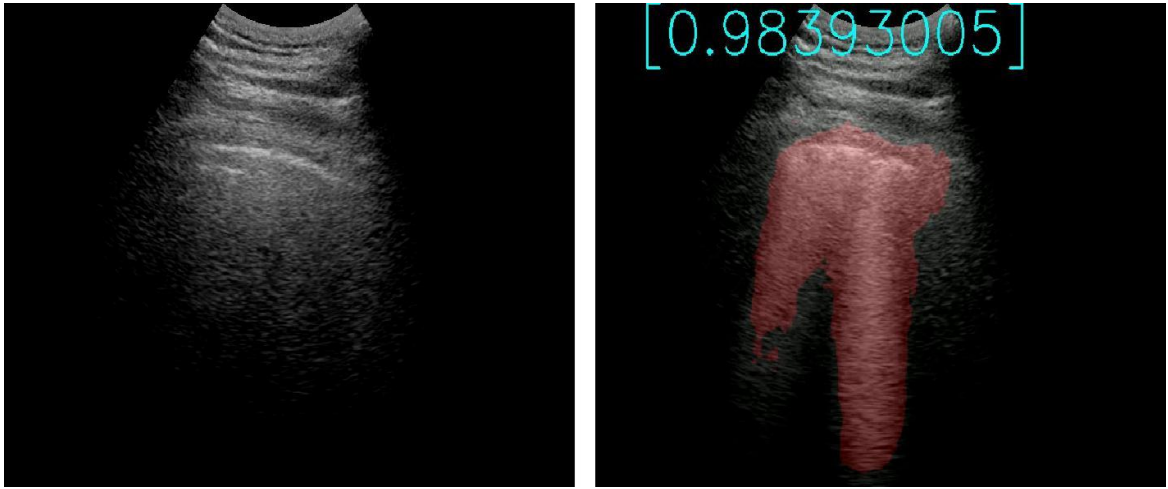

**Supplementary Figure 1: Preferential (diagnostically important) summarization aspect of the algorithm. A) Presence of ambiguous regions in the ultrasound scan and B) the preferential selection of pathologically relevant structures like the B-lines for the summary.**

## Web-application Deployment for Tele-medicine

To further validate the usefulness of the proposed methodology in tele-medicine, a prototype of the system is developed into a web application as a proof of concept for deployment on cloud. The demo of the tool can be found in [2] and is available for trial by contacting the authors. A sample web-application snapshot is shown in Supplementary Fig.2.

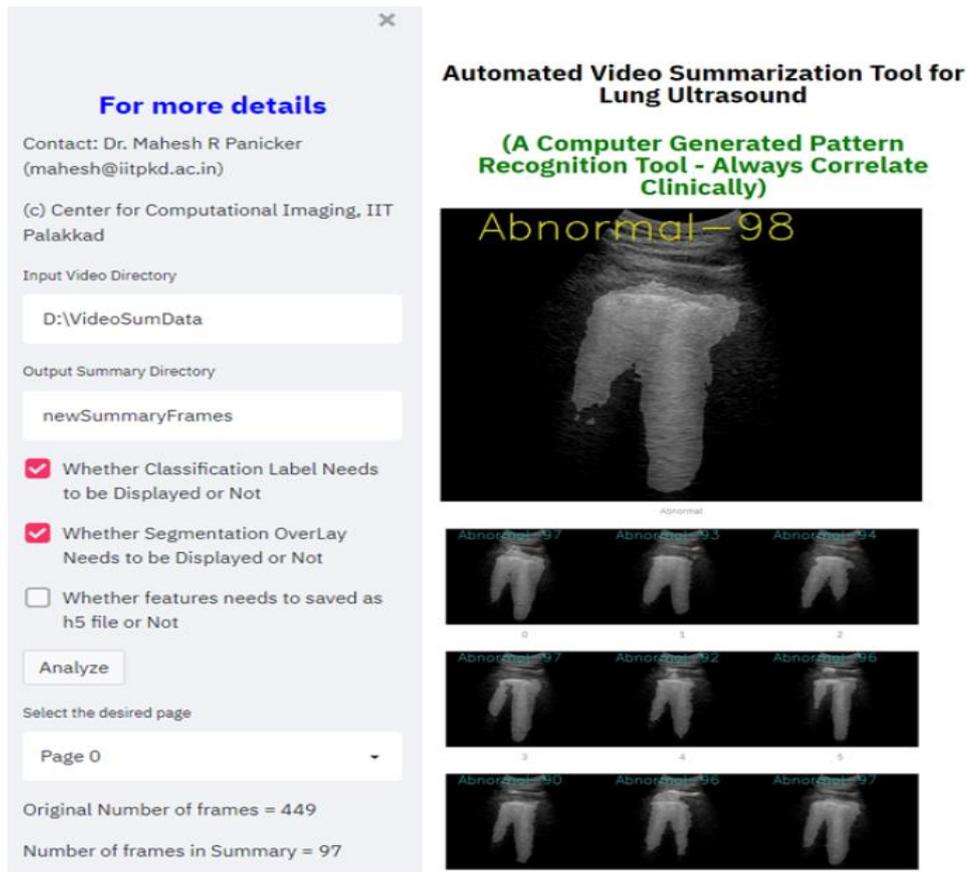

**Supplementary Figure 2:** Web-application deployment. The tool has the option to select the classification labels and the segmentation maps to be overlaid on the frames. Provision is also given to save the frame features. The tool will show the summary video, a collage of random 9 frames from the summary video and the frame numbers.

**Comments and Disclaimer:**

1. As mentioned in the main text, the present work demonstrates a prototype for a robust ultrasound video summarization pipeline. At present the system is trained and validated with ultrasound scans from various geographies obtained from different ultrasound machines by separate clinicians. Future work would include analyzing the proposed system with better pruned US scans i.e., data from distinct machine vendors, standardized ultrasound scans with specific preset which are expected to further increase the performance of the proposed methodology.

2. The following prototype is in development and should not be used as a substitute for a trained professional in making diagnostic decisions.

**Supplementary References**